\documentclass[acmsmall,screen]{acmart}

\setcopyright{none}
\copyrightyear{2025}
\acmYear{2025}
\acmDOI{}

\usepackage{fancyhdr}
\usepackage{ tipa }

\DeclareUnicodeCharacter{2080}{$_{0}$}
\DeclareUnicodeCharacter{2081}{$_{1}$}
\DeclareUnicodeCharacter{2082}{$_{2}$}
\DeclareUnicodeCharacter{2083}{$_{3}$}
\DeclareUnicodeCharacter{2084}{$_{4}$}
\DeclareUnicodeCharacter{2085}{$_{5}$}
\DeclareUnicodeCharacter{2086}{$_{6}$}
\DeclareUnicodeCharacter{2087}{$_{7}$}
\DeclareUnicodeCharacter{2088}{$_{8}$}
\DeclareUnicodeCharacter{2089}{$_{9}$}
\DeclareUnicodeCharacter{03BB}{\ensuremath{\lambda}}
\DeclareUnicodeCharacter{3B4}{$\bm{\delta}$}
\DeclareUnicodeCharacter{3B8}{$\bm{\theta}$}
\DeclareUnicodeCharacter{3BB}{$\lambda$}
\DeclareUnicodeCharacter{3BC}{$\mu$}
\DeclareUnicodeCharacter{0308}{\"o}
\usepackage{pmboxdraw} %% Handles unicode, including boxes

\usepackage{amsmath}
\usepackage{graphicx}
\usepackage{subcaption}
\usepackage{minted}
\newminted[code]{haskell}{autogobble,frame=lines}
\newminted[spec]{haskell}{autogobble,frame=lines}
\newminted{cpp}{gobble=2,linenos}
\newcommand{\il}[1]{\mintinline{haskell}{#1}}

\newcommand{\ToolName}{\textsc{Walrus}}
\newcommand{\OurTool}{\ToolName}

\newcommand{\savecolumns}{}

\usepackage{lipsum}

\usepackage{verbatim}

\usepackage[textsize=tiny]{todonotes}
\begin{document}

\title{Designing \ToolName{}: Relational Programming with Rich Types,
On-Demand Laziness, and Structured Traces}
\author{Santiago Cuellar   }
\affiliation{%
  \institution{Galois, Inc.}
  \city{Arlington, VA}
  \country{USA}
}
\email{santiago@galois.com}

\author{Naomi Spargo      }
\affiliation{%
  \institution{Galois, Inc.}
  \city{Arlington, VA}
  \country{USA}
}
\email{nspargo@galois.com}

\author{Jonathan Daugherty}
\affiliation{%
  \institution{Galois, Inc.}
  \city{Portland, OR}
  \country{USA}
}
\email{jtd@galois.com}

\author{David Darais      }
\affiliation{%
  \institution{Galois, Inc.}
  \city{Portland, OR}
  \country{USA}
}
\email{darais@galois.com}
\thanks{This material is
based upon work supported by the Defense Advanced Research Projects
Agency (DARPA) and Naval Information Warfare Center Pacific (NIWC
Pacific) under Contract No. N66001-21-C-4023.  Any opinions, findings
and conclusions or recommendations expressed in this material are
those of the author(s) and do not necessarily reflect the views of
DARPA and NIWC Pacific.}

%%%%%%%%%% == FORMATTING INSTRUCTIONS (From miniKanren workshop) ==
\settopmatter{printacmref=false}
\settopmatter{printfolios=true}
\renewcommand\footnotetextcopyrightpermission[1]{}
\pagestyle{fancy}
\fancyfoot{}
\fancyfoot[R]{miniKanren'25}
\fancypagestyle{firstfancy}{
\fancyhead{}
\fancyhead[R]{miniKanren'25}
\fancyfoot{}
}
\makeatletter
\let\@authorsaddresses\@empty
\makeatother
%%%%%%%%%% ============================

\begin{abstract}

\begin{ignore}
\begin{code}
module Abstract where
\end{code}
\end{ignore}

We present \ToolName{}, a functional relational programming language
embedded in Haskell that extends the miniKanren model with
type-polymorphic unification, on-demand laziness, and a range of
usability features aimed at practical development. These include
use of Haskell Generics for boilerplate reduction, structured debugging traces,
and ergonomic support for product types. We describe the design and
implementation of \ToolName{} through the lens of our experience
developing bidirectional compilers, and reflect on key design decisions
and recurring usability challenges encountered in practice.

\end{abstract}

\maketitle

%%%%%%%%%% == FORMATTING INSTRUCTIONS (From miniKanren workshop) ==
\thispagestyle{firstfancy}
%%%%%%%%%% ============================

\begin{ignore}
\begin{code}
module Main where
import Walrus

main :: IO ()
main = putStrLn "Hello World!"
\end{code}
\end{ignore}

\section{Introduction}
\label{sec:intro}
\begin{ignore}
\begin{code}
module Introduction where
\end{code}
\end{ignore}

Relational programming is a programming paradigm where computations are defined
as relations instead of functions.
Whereas functions encode one-way mappings between inputs and outputs, relations
encode bidirectional constraints between values without a priori distinctions
of ``input'' versus ``output''.
As a paradigm, relational programming refines the more general concept of logic
programming with the restriction that all relational constructs must be
``pure''.
When computing with relational programs, arguments may remain unspecified or
underspecified, and evaluation proceeds by solving for values that satisfy all
of the imposed constraints.

One use case for relational programming is the ability to run a computation
``backwards''~\cite{miniKanren-quines}.
Byrd et al demonstrated a powerful instantiation of this idea in the context of
programming language interpreters with Barliman~\cite{barliman}, a Scheme
interpreter implemented in miniKanren, a relational programming language.
The result is an interpreter \emph{relation} between Scheme programs and their
input/output behavior. Because the interpreter is implemented in miniKanren, it
``inherits'' the features of relational programming and can be run backwards
to synthesize programs from input/output examples and other partial
specifications.

In this paper we present \ToolName{}, a functional logic programming language
based on miniKanren and embedded in Haskell.
Our goal in building \ToolName{} was similar to that of Barliman: to support
the development of relational programming language \emph{compilers}---rather
than \emph{interpreters}---so they can be run ``backwards'' to achieve
decompilation.
Our results show that the relational programming ideas behind Barliman can be
transferred successfully to the domain of relational compilers, exploiting the
deep structural similarity between interpreter and compiler domains.
We describe the design and implementation of \ToolName{} in this paper from a
general-purpose relational programming perspective, and defer the description
of how \ToolName{} is instantiated to achieve decompilation (lifting C code to
a high level DSL) to a separate paper~\cite{cocompiler}.

Many implementations of logic programming already exist, and especially so for
variants of miniKanren and microKanren, due largely to their beauty and
simplicity.
However, each implementation comes with fundamental trade-offs in their design
that bias their performance and usability towards certain use cases.
After surveying existing implementations, we decided to roll our own.
We made this choice to have more control over performance and user-imposed
boilerplate specifically for our intended use case of relational compilers.

The design of \ToolName{} is closest to that of
faster-miniKanren~\cite{faster-miniKanren} and typedKanren~\cite{typedKanren},
however it also offers some new innovations not found in either (or any, to our
knowledge) prior system as detailed below.

By way of technical innovations, we contribute the following novel techniques
in the design and implementation of relational programming languages:
\begin{itemize}
\item
\textbf{A type-segregated substitution store.}
Substitutions are indexed by the concrete Haskell type of each logic variable,
avoiding unsafe casts and improving lookup performance.
\item
\textbf{Lazy relational evaluation with disequality.}
Laziness allows some goals to be evaluated on demand, reducing state explosions
and expensive lookups through a single general-purpose language construct.
\item
\textbf{Usability features.}
These include ordered debugging traces, multiple strategies to eliminate
boilerplate, and first-class support for product types. Together, they lower
the barrier to applying logic programming inside Haskell.
\end{itemize}

In terms of performance, we observe that \ToolName{} is slower than
faster-miniKanren and comparable to typedKanren (both faster and slower
depending on the benchmark).
We speculate that faster-miniKanren is faster than \ToolName{} primarily due to
general tool maturity of faster-miniKanren, and performance differences between
\ToolName{} and typedKanren are primarily due to different approaches for
embedding in Haskell's type system (see Section~\ref{sec:substitutions}).
To arrive at these performance assessments we compared the runtime performance
of each implementation on 5 benchmark programs: 2 programs that perform
arithmetic calculations over binary encodings (exponent and logarithm), and 3
variants of quine generation.

The remainder of this paper is structured as follows.
Section~\ref{sec:example} introduces \ToolName{} by working through the example
of generating quines.
Section~\ref{sec:related-work} discusses relevant background and related work.
Section~\ref{sec:foundations} presents the core concepts and design of
\ToolName{}.
Section~\ref{sec:generics} describes our use of Haskell Generics to reduce
boilerplate.
Section~\ref{sec:lazy} describes our implementation of laziness and disequality
constraints.
Section~\ref{sec:productTypes} describes our approach to reducing boilerplate
for user-defined product types.
Section~\ref{sec:debug} describes our approach to structured debugging and
debug trace generation.
Finally, Section~\ref{sec:evaluation} presents an evaluation,
Section~\ref{sec:future-work} outlines future directions, and
Section~\ref{sec:conclusion} concludes.

This paper is written entirely in literate Haskell and is executable. For
brevity, some pieces of code are not rendered in the pdf version but are be
available in the source \il{.lhs} files.

\section{Introductory Example}
\label{sec:example}
\begin{ignore}
\begin{code}
{-# OPTIONS_GHC -w #-}
{-# LANGUAGE FlexibleInstances #-} -- REmove after isString is moved
{-# LANGUAGE TypeApplications #-}
{-# LANGUAGE OverloadedStrings #-}
{-# LANGUAGE DeriveGeneric #-}
module Example where
\end{code}
\end{ignore}

To showcase \ToolName{} in action, we begin with an overview of a
classic example: quine generation. Quines~\cite{hofstadter}, programs
that output their own source, are a well-known benchmark for
miniKanren implementations~\cite{hofstadter}. In \ToolName{}, they
offer an expressive demonstration of how logical unification
integrates with Haskell types. To generate quines using relational
programming, one first implements an interpreter for a core lambda
calculus extended with support for quoted program terms. Then this
interpreter is instantiated with the same unification variable for the
input and output, which generates programs that evaluate to
themselves.

\begin{ignore}
\begin{code}
import           Walrus hiding (VStr)
import           Prettyprinter
import           GHC.Generics (Generic)

-- Move this to the core library
import           Data.String
import Data.Kind (Type)

instance IsString (UT String) where
  fromString = UT
\end{code}
\end{ignore}

We follow with a minimal interpreter for a simple lambda calculus. Its
core data structures\hspace{0pt}---\hspace{0pt}environments, terms, and closures—show how
ordinary Haskell types can be lifted into \ToolName{} and enriched
with logical variables for symbolic evaluation.

To begin, the environment is a list of (name, term) pairs, represented
using \ToolName{}’s list type \il{L} type. We lift standard Haskell
tuples and strings into logical types with the \il{U} and \il{UT}
constructors (for unifiable and unifiable terminal respectively),
which add logical variables to a type and help \OurTool{} understand
how to unify it, as explained in Section~\ref{sec:logical-vars}. The
type operators~\il{L},~\il{U}, and~\il{UT} are of kind~\il{ Type ->
Type}.

\begin{code}
type VStr = UT String -- Part of the Walrus standard library
type Env = L (U (VStr, Lam))
\end{code}

We define a data type for lambda terms, including variables,
abstractions, and applications, quoted terms, a quoting operator, and
pairs.

\begin{code}
data Lam = LVar VStr | LLam VStr Lam | LApp Lam Lam
         | LQuoted Lam | LQuote | LPair Lam Lam | LUVar ID
         deriving (Eq, Show, Generic)
\end{code}

Notice the last constructor, \il{LUVar ID}, which represents a logical
variable and turns \il{Lam} into a logical type. Since this is a
custom definition, we need to tell \ToolName{} which constructor
represents the logical variables of this type. Such an \il{ID}-carrying
constructor is a required step in equipping most Haskell types for use in
\ToolName{}. Next, we provide a \il{Unifiable} type class instance for the
type. The \il{uvar} operation injects a unification variable into the
user-defined type, and the \il{isUVar} operation extracts a unification
variable from the user-defined type. That instance, explained in
Section~\ref{sec:unification}, prompts the compiler to generate all the
necessary boilerplate for unification.

\begin{code}
instance Unifiable Lam where
  uvar i = LUVar i
  isUVar (LUVar i) = Just i
  isUVar _ = Nothing
\end{code}

The interpreter's return type \il{Value}, which includes closures,
does not contain logical variables. We can still use it as a logical
type using \il{U} as long as we add a \il{PreUnifiable} instance for
\il{Value}. As explained in Section~\ref{sec:unification}, this empty
declaration automatically generates the code needed to make \il{U
Value} into a logical type.

\begin{code}
data Value = VL Lam | Closure VStr Lam Env
  deriving (Eq, Generic)
instance PreUnifiable Value where
\end{code}

\begin{ignore}
\begin{code}
instance Pretty Lam where
  pretty l =
    case l of
      LUVar i -> "\\"<> pretty i
      LVar name -> pretty name
      LLam name lam -> "λ" <> pretty name <> "." <+> pretty lam
      LApp l1 l2 -> parens (pretty l1) <+> parens (pretty l2)
      LQuoted lam -> ("\"" <> pretty lam <> "\"")
      LQuote  -> "quote"
      LPair l1 l2 -> parens (pretty l1 <> ", " <>pretty l2)

instance Pretty Value where
  pretty (Closure str lam env) =
    "[clo " <> ( pretty str <> " , " <> pretty lam  <> " | " <>  pretty env) <> "]"
  pretty (VL lam) = pretty lam

\end{code}
\end{ignore}

To interpret lambda terms in \ToolName{}, we define the relation
\il{interp} stating that in environment \il{env} the expression
\il{exp} evaluates to \il{val} in big-step semantics. The relation is
defined as a disjunction of cases corresponding to a pattern match on
\il{exp}, or equivalently, on \il{val}. In \ToolName{}, this
disjunction is expressed using \il{disjP} which plays the role of
\il{disj+} in \texttt{miniKanren}. Conjunction is written using
Haskell’s \il{do} notation, aligning with the monadic structure of
goals, the computation type of \ToolName~(Section~\ref{sec:goal}).
Readers familiar with \texttt{miniKanren} will recognize this pattern
as directly corresponding to \il{conde}. For brevity, we display only
the most illustrative cases\footnote{The entire example is present and
executable in the literate Haskell verions of this paper.}.

%If this casues an error, you are likely using an older version of minted. See README
\begin{code}
interp :: Lam -> Env -> U Value -> Goal ()
interp exp env val = disjP $
    [ do str <- fresh
         lam <- isLambda val
         LVar str === exp
         searchEnv env str lam
    , do (e1, e2) <- freshProd
         LApp e1 e2 === exp
         (str, body, env', a) <- freshProd
         interp e1 env (U (Closure str body env'))
         interp e2 env (U (VL a))
         interp body (Cons (U (str, a)) env') val
    , do (str, body) <- freshProd
         LLam str body === exp
         U (Closure str body env) === val
     -- Some cases omitted
    ]
\end{code}
\begin{ignore}
\begin{code}
    ++ [
    do lam <- isLambda val
       LQuoted lam === exp
    , do (a, e2) <- freshProd
         LApp LQuote e2 === exp
         lam <- isLambda val
         lam === LQuoted a
         interp e2 env (U (VL a))
    , do (e1, e2) <- freshProd
         LPair e1 e2 === exp
         (a1,a2) <- freshProd
         lam <- isLambda val
         lam === LApp a1 a2
         interp e1 env (U (VL a1))
         interp e2 env (U (VL a2))
    ]
\end{code}
\end{ignore}

In the case involving \il{LVar}, we assert that \il{val} is a lambda term
rather than a closure. This is expressed using the monadic function
\il{isLambda}, which both enforces that \il{val} has the
expected structure and \emph{returns} the underlying lambda term. This
pattern highlights a key feature of \ToolName{}: its dual nature as both a
relational and functional language.

\begin{code}
isLambda :: U Value -> Goal Lam
isLambda val = do lam <- fresh
                  val === U (VL lam)
                  return lam
\end{code}

\begin{ignore}
\begin{code}

searchEnv :: Env -> VStr -> Lam -> Goal ()
searchEnv env str val = -- Implementation in Section~\ref{sec:lazy}
  disj
    [ do env' <- fresh
         Cons (U (str, val)) env' === env
    , do (x, val', env') <- freshProd
         Cons (U (x, val')) env' === env
         searchEnv env' str val
    ]
\end{code}
\end{ignore}

To evaluate the \il{interp} relation, we can use
\ToolName{}'s function \il{run} which takes two arguments: the number
of expected solutions \il{n} and a predicate of type \il{t -> Goal()}.
Similar to \texttt{miniKanren}'s \il{run} method, this execution
returns at most \il{n} values of type \il{t} that satisfy the
predicate. To generate a quine we run the interpreter on an empty
environment and requiring the output to match the input term. The
result is rendered using a pretty printer\footnote{Implementation of
the pretty printers are omitted in this paper.}:

\begin{spec}
λ>  pretty $ run 1 (\q -> interp q Empty (U (VL q)))
[(λ_₃. (_₃, (quote) (_₃))) ("λ_₃. (_₃, (quote) (_₃))")]
\end{spec}

The small but rich example in this section showcases the key
capabilities of \ToolName{}, including creating logical types in
Haskell, writting relational programs, and using \ToolName{} as
a functional language. In the following sections, we explore the
language’s core design and implementation choices in greater detail.

\begin{ignore}
\begin{code}
findQuine :: Lam -> Goal ()
findQuine lam = do
  interp lam Empty (U (VL lam))

findTwine :: U(Lam,Lam) -> Goal ()
findTwine lams = do
  (l1,l2) <- freshProd
  lams === U (l1, l2)
  interp l1 Empty (U (VL l2))
  interp l2 Empty (U (VL l1))

quine1 :: Lam
quine1 = LApp lhs (LQuoted lhs)
  where
    lhs = LLam "x" $ LPair (LVar "x") quotedX
    quotedX = LApp LQuote lhs


\end{code}
\end{ignore}

\begin{ignore}

Lifting quine:

(λ_₃. (_₃, (quote) (_₃))) ("λ_₃. (_₃, (quote) (_₃))")

let foo x = (_₃, (quote) (_₃)) in foo ("λ_₃. (_₃, (quote) (_₃))"

\end{ignore}

\section{Background and Related Work}
\label{sec:related-work}

\begin{ignore}
\begin{code}
module RelatedWork where
\end{code}
\end{ignore}

One of the distinguishing features of the \texttt{miniKanren} family is its
diversity: the core ideas have been adapted across a wide range of
host languages, design goals, and implementation strategies. Among
these, we focus on comparisons with typed embeddings of \texttt{miniKanren},
particularly those in functional languages where the challenges of
integrating logic programming with static type systems are directly
addressed.

Numerous \texttt{miniKanren} variants have been implemented in both
Haskell and OCaml~\cite{miniKanren.org}, yet most are monomorphic with
a single logical type. Early Haskell systems such as \texttt{Molog}
\cite{Molog} and \texttt{miniKanrenT} \cite{miniKanrenT} introduced
polymorphic unification, but doing so required a large amount of
boilerplate. Around the same period, the unpublished project
\texttt{HKanren}\cite{hkanren} used a clever functor-based encoding
that eliminated most boilerplate for logical types. However, this
convenience comes at the cost of a considerably more intricate type
class hierarchy and a heavier reliance on advanced type-system
features. Although we suspect that a simpler core could be distilled
from the approach or cleverly hidden behind a clean interface, the
repository has remained dormant for nearly a decade.

Several Haskell implementations, including the three discussed above,
rely directly on the \il{LogicT} monad
\cite{LogicT-repo,backtracking} or an equivalent
construction. \il{LogicT} manages interleavings and backtracking
without explicit suspensions, an approach that contrasts with the
suspended-stream semantics presented in
\texttt{miniKanren}~\cite{reasoned-schemer} and adopted in \ToolName{}
(cf. Section~\ref{sec:stream}).

\texttt{OCanren} \cite{ocanren} tackles boilerplate in OCaml by
exploiting the language’s uniform runtime representation and carefully
controlled type casts. This allows fully polymorphic unification
without requiring per-type implementation. The trade-off is that
unification must follow runtime tags rather than compile-time types,
which weakens the abstraction benefits of a typed interface (see
Section~\ref{sec:unification}). Reliance on low-level casts can also
complicate maintenance and conceal subtle errors. Although the authors
cite the continuation-based \il{LogicT} monad, the implementation is
actually a suspended stream in the style of the origianl
\texttt{miniKanren}.

A similar design is adopted by \texttt{klogic}~\cite{klogic}, a typed
\texttt{miniKanren} embedding for Kotlin on the JVM, which follows
OCanren’s overall strategy of polymorphic unification based on runtime
reflection and suspended‐stream search.

\texttt{typedKanren} \cite{typedKanren} avoids boilerplate by deriving
polymorphic unification from Haskell’s \il{Generic} representations
(Section~\ref{sec:generics}). Its search strategy is based on a custom
suspended-stream monad, not \il{LogicT}, and it relies on Haskell’s
laziness to avoid explicit thunks or explicit forcing
mechanisms. \ToolName{} was developed independently, yet it converges
on many of the same design choices, which we view as corroboration of
a sensible common core. \texttt{typedKanren} goes beyond our current
implementation by using metaprogramming to lift arbitrary data types
into their logical counterparts automatically. Although we have not
yet needed such automation, it will likely prove essential for a fully
general typed \texttt{miniKanren} in Haskell. \texttt{typedKanren}
also introduces a \il{match}\textsuperscript{\il{e}} construct that
performs exhaustiveness checking when pattern-matching on sum
types. Our work in Section~\ref{sec:productTypes} provides a
complementary facility for product types.

Execution order is a well-studied concern in \texttt{miniKanren} and,
more broadly, in relational programming. Implementations employ a
variety of strategies, from fair interleaving to explicit delay
combinators. \ToolName{} provides lazy evaluation on demand: the
programmer can explicitly delay the execution of a predicate until its
arguments are concrete (i.e., fully instantiated). This mechanism is
used selectively when deferred search is beneficial, while the default
execution strategy remains eager. This mechanism generalises the
optimization that \texttt{fasterMiniKanren} applies to disequality
constraints and, to the best of our knowledge, represents the most
flexible formulation of controlled laziness to date.

\section{Core Concepts and Design}
\label{sec:foundations}
\begin{ignore}
\begin{code}
{-# LANGUAGE DeriveGeneric #-}
{-# LANGUAGE FlexibleContexts #-}
{-# LANGUAGE DefaultSignatures #-}
{-# LANGUAGE KindSignatures #-}
{-# LANGUAGE TypeFamilies #-}
module Foundations where

import           Walrus hiding (Unifiable(..), ID)
import           Data.Map (Map)
import qualified Data.Map as Map
import           Data.TypeMap.Dynamic (TypeMap, Item)
import           Data.Kind (Type)
import           Data.Typeable (Typeable)
import           Prettyprinter
import           GHC.Generics hiding (S)
import           Control.Applicative (Alternative(..))
import           Control.Monad.State (MonadState(..))
import           Walrus.Debug (DebugM)
import qualified Walrus.Debug as Dbg
import           Walrus.Lib.Arith.Integers (mul)
import           Control.Monad.State (StateT)
import           Walrus.Lib.Arith.BinaryNumbers (Z)

\end{code}
\end{ignore}

At the core of \ToolName{} you will find the familiar ingredients of many
miniKanren implementations:

\begin{enumerate}

\item \emph{Goals}, the fundamental unit of computation in a
      relational program. A goal represents a constraint or relation
      to be satisfied and may succeed, fail, or produce multiple
      solutions when evaluated.

\item \emph{Logical variables}, which represent unknown values to be
      instantiated through unification.

\item \emph{Substitutions}, mappings from logical variables to terms,
      are used to record and propagate variable bindings as constraints
      are resolved.

\item \emph{Unification}, the core operation of logic programming,
      which determines whether two terms can be equated under a
      given substitution.

\item \emph{Streams and search strategy}, which represent the evolving
      search space of a program by tracking the state of computation
      across multiple branches. The search strategy ensures fairness
      by visiting all branches periodically.

\end{enumerate}

The rest of this section describes how to implement these elements
in Haskell to support polymorphic unification with type-segregated
lazy-substitution.

\subsection{The \il{Goal} monad}
\label{sec:goal}

Much of the complexity described in this section is encapsulated within
the \il{Goal} abstraction. \ToolName{} users are encouraged to build
computations using the high-level \il{Goal} combinators provided by
\ToolName{}. Even advanced users, including contributors to the standard
library, rarely need to look beyond the core abstractions described
in this section. The internal structure shown here is not part of the
public API and is deliberately kept hidden to maintain a clean and
consistent programming interface. This confers the standard benefits of
abstraction: maintaining key invariants, ensuring that the API is not
error-prone to use, and ensuring that no knowledge of the internals is
needed for users to be successful with \ToolName{} programming.

Internally, the \il{Goal} monad coordinates several responsibilities
critical to \ToolName{}’s logic engine:

\begin{itemize}

\item It maintains local state, including substitutions and fresh
      variable generation, using the state monad \il{StateT} and
      \ToolName{}'s state \il{WState}.

\item It manages the search space and ensures fair backtracking via
      the \il{SuspLogicT} monad, described in Section~\ref{sec:stream}.

\item It tracks global metadata for debugging, such as branching
      history and evaluation traces, using the debugging
      infrastructure described in Section~\ref{sec:debug}.

\end{itemize}

The underlying structure is made explicit in
the implementation of the \il{Goal} monad itself:

\begin{spec}
data WState =
  WState { nextVarW :: ID        -- ID management
         , subsW :: AllSubs'     -- Substitutions
         , dbgPathW :: Dbg.Path  -- Local debug info
         }

type InGoal a = StateT WState (SuspLogicT DebugM) a

data Goal a = MkGoal { unGoal :: InGoal a }
\end{spec}

\il{Goal} is polymorphic in a return type and allows \ToolName{} to
function as a purely functional language in addition to being a
relational one. We saw this in Section~\ref{sec:example}, where the
\il{isLambda} predicate also returns the internal structure of its
argument when it succeeds. Below, we show both the relational and
functional versions of a simple \il{square} computation, where \il{Z}
is the logical type for integers, as defined in \OurTool's standard
library. While this does not increase the expressive power of the
language, since any function can be encoded as a relation and vice
versa (as shown by \il{squareF}), it significantly improves ergonomics
and modularity. In particular, functions sometimes eliminate the need
to introduce auxiliary unification variables when extracting values
from relational contexts.

\begin{code}
-- | Square as a relation.
squareR :: Z -> Z -> Goal ()
squareR rt sq = mul rt rt sq

-- | Square as a function
squareF :: Z -> Goal Z
squareF rt = do
  sq <- fresh
  mul rt rt sq
  return sq
\end{code}

With that said, writing \ToolName{} code in a functional style comes with
trade-offs. First, functions cannot be run ``backwards'' like relations,
limiting their use in bidirectional or generative contexts. Second,
functionally-structured programs often fail later in the search
process than relationally-structured ones, which can negatively impact
performance. Despite these limitations, we frequently use functional
style for convenience, including for internal combinators, such as
\il{fresh}.

It is important to note that, unlike most monads, the \il{Goal} monad
is not associative in the usual sense---it is only associative up to
reordering of results. This is due to the way computations interleave:
re-associating \il{Goal} expressions can change the order in which
solutions are produced. Even seemingly harmless transformations, such
as function inlining\footnote{While such a transformation would not be
performed by the GHC compiler, it remains a reasonable and natural
refactoring for a developer.}, may shift result order.

This behavior is closely related to the well-known difficulties in
defining a correct \il{ListT} monad
transformer~\cite{ListT-done-right}, where interleaving and fairness
complicate standard monadic laws. However, because our interleaving
strategy is fair~\cite{backtracking} (Section~\ref{sec:stream}), the
set of results remains the same even though the order may differ. In this
sense, we say that \il{Goal} satisfies the monad laws up to
reordering, and behaves correctly when results are interpreted as sets
rather than sequences.

This subtlety has significant performance implications. In relational
expressions such as \newline \il{producer >= } \il{filter} the order in which
\il{producer} generates values can affect how quickly \il{filter}
encounters an acceptable input. If admissible values are delayed, the
overall computation may take significantly longer to produce a single
result. Thus, while the set of possible results remains the same due
to fair interleaving, the overall responsiveness can vary greatly
based on ordering.

Despite this non-associativity, we find that embracing the monadic
interface, including \il{do}-notation and monad combinators, greatly
enhances usability. In practice, the benefits in clarity and
composability outweigh the potential confusion introduced by
non-deterministic execution order.

\subsection{Logical variables and logical types.}
\label{sec:logical-vars}

\ToolName{} represents logical variables using the opaque identifier type
\il{ID}. Fresh variables can only be generated within a stateful
monad that tracks allocated identifiers (Section~\ref{sec:goal}).

\begin{code}
data ID = ID Int
\end{code}

To support polymorphic unification, all types in \ToolName{} must
contain logical variables. We refer to such types as \emph{logical
types} or \emph{\ToolName{} types}. Most types are made logical using
the type constructors \il{U} and \il{UT}, which provide a concise,
boilerplate-free mechanism to lift existing Haskell types into
\ToolName{}.

\begin{code}
data U t = UVar ID | U t
data UT t = UTVar ID | UT t
\end{code}

The key distinction between these two lies in how unification is performed:

\begin{itemize}

\item \il{UT} is designed for types that are treated as indivisible
during unification, which we call \emph{atomic} or
\emph{terminal}—hence the \texttt{T}. Concretely, \il{UT t} is unified
by structural equality in Haskell or by creating new associations, but
their internal structure is not inspected or decomposed. This makes
unification of \il{UT} types much faster and is appropriate for
types such as \il{Void}, \il{()}, \il{Bool}, \il{Ordering},
identifiers, or memory pointers that only support equality
comparison.

\item By contrast, types lifted with \il{U} are intended for
structured or composite types. These are values whose internal
components may themselves be subject to unification. Unification on
these types may descend recursively into their structure, unifying
subcomponents when necessary. The \ToolName{} standard library
includes \il{U}-based wrappers for common Haskell types such as
tuples, \il{Maybe}, and \il{Either}, making them readily usable as
logical types. In Section~\ref{sec:productTypes} we also show how to
lift Haskell functions like field accessors, for types wrapped with
\il{U}.

\end{itemize}

Alternatively, users may define custom types with logical variables
embedded directly in their constructors. For example, the
\emph{recursive} \il{Lam} type includes the constructor \il{LUVar} to
represent logical variables explicitly. In such cases, users must
implement the \il{uvar} and \il{isUVar} methods of the \il{Unifiable}
type class to specify how variables are constructed and identified. We
demonstrated this for \il{Lam} in Section~\ref{sec:example},
illustrating how straightforward it is in practice. The complete
\il{Unifiable} type class is described in Section~\ref{sec:unification}.

\subsection{Type-segregated substitutions}
\label{sec:substitutions}

To support polymorphic unification, \ToolName{} must track subsitutions for
all types. However, uniquely to \ToolName{}, we keep mappings for each type
separated. Concretely, for any type \il{t} we create an internal type
\il{Subst'} wrapping a map from variable identifiers to terms of
type \il{t}. To make \ToolName{} \emph{polymorphic}, we store these
substitutions\footnote{These definitions
are marked with a prime because they are later revised when we
introduce \il{thunk}s in \ref{sec:lazy}} in a \il{TypeMap}~\cite{type-map}. We explore the
performance benefits of this approach in Section~\ref{sec:evaluation}.

\begin{code}
data Subst' t = Subst' { getSubstMap :: Map ID t}

data AllSubs' = AllSubs' (TypeMap S)

data S

type instance Item S t = Subst' t
\end{code}

Here, \il{S} serves as a type-level key for \il{TypeMap},
allowing substitution maps to be indexed by type. The associated
\il{Item} type family specifies that each type \il{t} maps to
a \il{Subst' t} containing values of that type. To perform lookups in
the \il{TypeMap}, type information must be reified at
runtime. This is achieved by requiring an instance of
\il{Typeable}, which enables the Haskell compiler to generate the
necessary type representations automatically. Concretely, to look up a
\emph{typed} identifier in the substitutions, we define a function of type

\begin{spec}
lookupSub :: Typeable t => ID -> AllSubs' -> Maybe t
\end{spec}

\noindent which first retrieves the appropriate \il{Subst'} for the
given type and then looks up the corresponding term within it. In the
opposite direction, we define an insertion function that operates
within a monadic state (denoted by \il{MonadState}). It retrieves the
current substitution map, updates the relevant type-specific
\il{Subst'} with a new association, and stores the updated state. The
state monad is needed to support laziness, as described in
Section~\ref{sec:lazy}.

\begin{spec}
insertSubM :: (MonadState AllSubs' m, Typeable t) => ID -> t -> m ()
\end{spec}

These two functions, \il{lookupSub} and \il{insertSubM}, form the entire
public interface for interacting with substitutions. The underlying
structure of substitutions is kept abstract, ensuring that users
manipulate them only through these safe and type-directed operations.

\subsection{Polymorphic unification}
\label{sec:unification}

\ToolName{} supports polymorphic unification over both user-defined
and existing Haskell types, balancing customization with ease of
use. By leveraging Haskell’s \il{Generic} representations, \ToolName{}
can automatically derive unification behavior for custom types to
eliminate the need for boilerplate. Users can also define their own unification rules
when the default behavior is insufficient, such as when the
unifiability of a set should not depend on the order of its elements,
when reasoning about lists should be lazy (see
Section~\ref{sec:lazy}), or to avoid the overhead of mapping \il{to}
and \il{from} generics.

Concretely, unification in \ToolName{} is invoked by the operator

\begin{spec}
(===) :: (Pretty t, Unifiable t)
      => t -> t -> Goal ()
\end{spec}

\noindent and applies to any two terms whose type implements the
\il{Unifiable} class.\footnote{The \il{Prettyprint} constraint is
largely omitted from this paper. It is used for the debugging
infrastructure described in Section~\ref{sec:debug} and we expect to
replace it with \il{Show}, which is automatically derivable.} This
class is the interface for logical types that support unification, can
be \il{walk}ed recursively, and implements the \il{occurs} check.

\begin{ignore}
\begin{code}
class GUnifiable f where
  gWalk :: AllSubs' -> f a -> f t
  gUnify :: (MonadState AllSubs' m, Alternative m) => f a -> f a -> m ()
\end{code}
\end{ignore}

\savecolumns
\begin{code}
class Typeable t => Unifiable t where
  uvar :: ID -> t
  isUVar :: t -> Maybe ID
  walk :: AllSubs' -> t -> t
  occurs :: ID -> t -> Bool
  unify :: (MonadState AllSubs' m, Alternative m) => t -> t -> m ()
\end{code}

The \il{Typeable t} constraint is required to enable type-indexed
substitutions for \il{Unifiable} types as discussed in
Section~\ref{sec:substitutions}. The first two methods, \il{uvar} and
\il{isUVar}, identify a logical type as described in
Section~\ref{sec:logical-vars}. The methods \il{walk} and \il{occurs}
behave as in standard miniKanren implementations, performing
substitution traversal and the \il{occurs} check,
respectively.\footnote{See \cite{reasoned-schemer} for
an introduction to these concepts.}

The \il{unify} method differs slightly from the traditional approach
that returns a modified substitution (e.g., \il{unifyMiniKanren :: t -> t ->
AllSubs' -> Maybe AllSubs'}). Instead, \ToolName{} defines unification
in a stateful monad. The unification function makes necessary updates
to the substitution from the monadic state. This design enables
support for laziness as described in Section~\ref{sec:lazy}.

\ToolName{} provides mechanisms to reduce boilerplate in a flexible way,
offering a sliding scale from zero effort to full manual control. At
the simplest level, any type of the form \il{UT t} is \il{Unifiable},
provided that \il{t} is an instance of both \il{Eq}\footnote{Equality
can be derived automatically or defined manually.} and
\il{Typeable}. \ToolName{} supplies an instance for such types as shown
below. In this case, \il{t} is assumed to be atomic, so the
unification process does not recursively inspect its structure.

\begin{spec}
instance (Eq t, Typeable t) => Unifiable (UT t) where
  -- Implementation omitted
\end{spec}

Similarly, any type of the form \il{U t} is \il{Unifiable}, provided
that \il{t} has instances for \il{Typeable}, \il{Generic}, and
\il{Eq}. These requirements are typically satisfied by most
non-recursive algebraic data types, such as the pair \il{(VStr, Lam)}
lifted using \il{U} in the definition of \il{Env} in
Section~\ref{sec:example}.

Supporting this level of automation is more involved than in the
\il{UT} case, as it relies on \il{Generic} to extract structure and
automate traversal. The user must declare an \emph{empty} instance of
a helper class \il{PreUnifiable}, which the compiler then populates
automatically. Many common cases like tuples, are built in in the
\ToolName{} standard library. For instance:

\begin{spec}
instance (Unifiable a, Unifiable b) => PreUnifiable (a, b) where
-- No implementation needed!
\end{spec}

Beyond simple lifting, \ToolName{} also supports fully-custom types
that include logic variables explicitly-just like \il{Lam}. Even in
this case, the core unification logic can still be automatically
derived as long as the type implements \il{Generic}. To enable this,
\ToolName{} provides default implementations for the key unification
methods \il{unify}, \il{walk}, and \il{occurs} as described in
Section~\ref{sec:generics}. Concretely, the following \il{default}
implementations are added to the definitions of the \il{Unifiable}
class.

\begin{spec}
  default unify :: ( MonadState AllSubs' m
                   , Alternative m
                   , Eq t
                   , Generic t
                   , GUnifiable (Rep t))
                => t -> t -> m ()
  unify lhs rhs = -- Omitted
  default walk :: (Generic t, GUnifiable (Rep t)) => AllSubs' -> t -> t
  walk subs x = -- Omitted
  default occurs :: (Generic t, GUnifiable (Rep t)) => ID -> t -> Bool
  occurs xid y = -- Omitted
\end{spec}

In practice, this means that users writing \il{Unifiable} instances
need only specify how to recognize and construct logical variables
with \il{uvar} and \il{isUVar}. The remaining methods will default to
generic implementations. Alternatively, users can override any or all
of these methods to implement fully custom unification behavior.

\subsection{\il{Stream}s and search strategy}
\label{sec:stream}

Many \texttt{miniKanren} implementations use \emph{suspended}
streams~\cite{reasoned-schemer} which are possibly-infinite lists, with
thunks.  Since Haskell is a lazy language, it does not need
\emph{suspensions} to represent infinite streams; the thunks come
built-in! However, suspensions are also a crucial part of miniKanren's
search strategy. For example, the interleavings described in \emph{The
{Reasoned} {Schemer}}~\cite{reasoned-schemer} are alternated at
suspensions, and only then. Therefore, it is not enough to use a
simple backtracking monad such as \il{LogicT}~\cite{LogicT-repo} since such
a monad alternates all \emph{results} and doesn't use suspensions, at
least not out of the box.

We define our own monad for backtracking computations with
suspensions, based on the streams used in \texttt{miniKanren}.
This monad is integrated into the \il{Goal} monad described in
Section~\ref{sec:goal}.

\begin{code}
newtype SuspLogicT m a =
  SuspLogicT { runSuspLogicT :: m (Stream (SuspLogicT m a) a) }

data Stream k a =
  Done         -- ^ Stream is empty
  | Suspend k  -- ^ No observable effect, but stream continues
  | SCons a k  -- ^ Observable output and continuation
\end{code}

\section{Generics}
\label{sec:generics}
\begin{ignore}
\begin{code}
{-# LANGUAGE DeriveGeneric #-}
module Generics where

import           GHC.Generics hiding (S)
import           Walrus (ID)

\end{code}
\end{ignore}

Some relational language implementations~\cite{ocanren} achieve
polymorphic unification by inspecting values at runtime, using
reflection or low-level operations to traverse terms without requiring
user-defined unification logic. While these approaches can be made
type-safe with sufficient care, they rely on runtime representations
that are tightly coupled to the language implementation and harder to
extend or adapt.

Rather than using runtime approaches, \ToolName{} uses Haskell’s
compile-time generics~\cite{generics} to assist in automatically
supporting unification for new data types. This approach preserves
safety and modularity without depending on runtime representation.
Haskell’s generics provide a metaprogramming framework that represents
user-defined algebraic data types as combinations of a few higher-level
type constructors like sums and products. This lets us define
unification for those constructors and allows Haskell to translate any type to
and from the generic representation.

As demonstrated in Section~\ref{sec:example}, we leverage Haskell’s
support for generic programming to automatically generate instances
for classes such as \il{Unifiable} and \il{PreUnifiable}. This is
accomplished in two steps. First, we implement a companion class
\il{GUnifiable} with all the functionality we need to generate and
implement instances for all generic type constructors. Then, we use
the \il{default} method mechanism to implement the
desired method from the generic version. For example, a default
implementation of the \il{walk} function is written:

\begin{spec}
class Typeable t => Unifiable t where
  -- Type signature constraints when the default is applied
  default walk :: (Generic t, GUnifiable (Rep t)) => AllSubs' -> t -> t
  walk subs x =
    case isUVar x of
      Just xid -> maybe x (walk subs) $ lookupSub xid subs
      Nothing  -> to $ gWalk subs $ from x
 -- Other methods omitted
\end{spec}

Generics are used on the last line, which works by converting the
input value to its generic representation using \il{from}, applying
the generic method \il{gWalk}, and then converting the result back to
the original type using \il{to}. \il{gWalk} is a method in the
\il{GUnifiable} class which we have implemented for all possible
generic representations. This structure is a standard pattern in the
use of generics~\cite{generics}. We describe more about \il{Unifiable}
in Section~\ref{sec:unification}.

\section{Lazyness and Disequality}
\label{sec:lazy}
\begin{ignore}
\begin{code}
{-# LANGUAGE OverloadedStrings #-} -- REmove after isString is moved
{-# LANGUAGE FlexibleInstances #-}  -- REmove after isString is moved
{-# LANGUAGE TypeFamilies #-}
{-# LANGUAGE KindSignatures #-}
{-# LANGUAGE DeriveGeneric #-}
module Lazyness where

import Walrus    hiding (AllSubs, mapL)
import           Prettyprinter
import           GHC.Generics (Generic)
import           Data.Map (Map)
import qualified Data.Map as Map
import           Data.Kind (Type)

import           Data.TypeMap.Dynamic (TypeMap, Item)
import           Control.Monad.State (MonadState(..))
import           Data.Typeable (Typeable)
import           Walrus.Lang.Substitution (AllSubs)
import           Data.Proxy (Proxy(..))

import Walrus.Lib.Arith.BinaryNumbers
import Walrus.Lib.Arith.Positives

import Debug.Trace

-- Move this to the core library
import           Data.String

instance IsString (UT String) where
  fromString = UT

\end{code}
\end{ignore}

Logic programs that operate over recursive structures such as lists
can suffer from excessive branching and state explosion. In some
cases, this can be mitigated through careful refactoring and
deliberate control over execution order. However, such restructuring
is often convoluted and occasionally infeasible.

The following example illustrates a common pattern we encountered in our
development:

\begin{ignore}
\begin{code}
-- To use integers directly
instance Num Positive where
  fromInteger = fromInt . fromEnum
  (+)    = undefined
  (*)    = undefined
  abs    = undefined
  signum = undefined
  negate = undefined


instance Enum Positive where
  toEnum = fromInt
  fromEnum (PUVar _) = error "Tried transforming a unification variable to an Enum"
  fromEnum PH = 1
  fromEnum (P a p) = 2 * (fromEnum p) + boolToInt a
    where boolToInt a = case a of
                          UT True -> 1
                          UT False -> 0
                          _ -> error "Tried transforming something with unification variable to an Enum"


mapL :: (Unifiable a, Pretty a, Unifiable b, Pretty b, Eq a, Eq b)
     => (a -> b -> Goal ())
     -> L a
     -> L b
     -> Goal ()
mapL r la lb =
  disjP [ do la === Empty
             lb === Empty

        , do ha <- fresh
             ta <- fresh
             la === Cons ha ta
             hb <- fresh
             tb <- fresh
             lb === Cons hb tb

             r ha hb
             mapL r ta tb
       ]

\end{code}
\end{ignore}

\begin{code}
expensiveComputation :: Goal ()
expensiveComputation = fresh >>= mapL (add 1) (toL [2..10])

eagerMap :: L Positive -> Goal ()
eagerMap ls = do
  temp <- fresh
  mapL (add 1) ls temp
  expensiveComputation
  temp === toL [2..10]
\end{code}

The problem arises when the above predicate is evaluated with a fresh
\il{ls}. Then the call to \il{map (add 1)}, which attempts to add 1 to
each element in \il{ls}, will eagerly match over the list, creating
infinitely many branches, one for each possible list. Then, each
branch will execute the \il{expensiveComputation}. Indeed, if we run
\il{run 1 eagerMap}, it times-out after 10 minutes. On the other hand,
the same computation is instataneous (0.02 secs) if we make the
\il{mapL} application lazy.

\begin{ignore}
\begin{code}
lazyMap ls = do
  temp <- fresh
\end{code}
\end{ignore}
\begin{code}
  mapL (add 1) ls `lazy` temp
\end{code}
\begin{ignore}
\begin{code}
  expensiveComputation
  temp === toL [2..10]
\end{code}
\end{ignore}

The dramatic difference in behavior arises from the fact that
\il{lazy} defers the evaluation of \il{mapL (add 1) ls} until
\il{temp} is unified.  This mechanism is inspired by Haskell’s
evaluation model in which expressions are lazy by default and are only
evaluated when their results are demanded. For example, the equivalent
Haskell expression does not traverse the list \il{ls} until \il{temp}
is \emph{used}:

\begin{spec}
temp = map (+1) ls
\end{spec}

\ToolName{} provides similar functionality through the \il{lazy}
combinator: a logic goal is suspended and only executed once its
associated variable is \emph{unified}. This allows precise,
demand-driven control over evaluation, which is especially useful when
working with recursive or nondeterministic structures. The next
subsection describes the implementation of \il{lazy} and the broader
design considerations for supporting laziness in a typed logic
language.  We then show how this mechanism can be used to support
disequality with little extra work.

While powerful, this feature, as implemented generically in \ToolName{},
has subtle drawbacks. First, it can delay failure, prolonging
computation along an ultimately failing branch. For instance, any
program invoking \il{lazyFail} below should never succeed, but will
continue evaluating until its argument becomes concrete:

\begin{ignore}
\begin{code}
gFail = gF
\end{code}
\end{ignore}

\begin{code}
lazyFail :: VStr -> Goal ()
lazyFail errorCode = (\_ -> gFail) `lazy` errorCode
\end{code}

Worse still, if the argument to \il{lazyFail} is never instantiated,
the associated thunk remains unforced, potentially resulting in an
inconsistent computation state. These issues become particularly
insidious when multiple, related thunks are involved. We have ideas
for mitigating these behaviors and plan to investigate them in future
work. For now, we leave readers with the guidance of a renowned New York
philosopher: With great power comes great responsibility.

\subsection{Design and Integration of Lazy Computation}

A delayed goal, or \emph{thunk}, is a computation of type \il{t -> Goal
()} associated with an uninstantiated logical variable of type
\il{t}. The intended behavior is that once the variable is unified with
a concrete value, the corresponding thunk is executed with that
value. Supporting thunks in \ToolName{} thus requires three ingredients:

\begin{enumerate}
\item Extending substitutions to store thunks.
\item Modifying unification to trigger thunks upon variable instantiation.
\item Providing an interface to register new thunks.
\end{enumerate}

\paragraph{Extending substitutions.} We enhance the type of substitutions \il{Subst}, defined in
Section~\ref{sec:substitutions}, so that variables can point to
either values or thunks, but not both. This ensures that only
unassociated variables can be related to thunks. Substitutions are
defined generically over any type of thunk and then instantiated with
\il{Goal}.

\begin{spec}
data Subst thunk t = Subst { getSubstMap :: Map ID (Either t thunk) }

data AllSubs thunk = AllSubs (TypeMap (S thunk))

data S (thunk :: Type -> Type)

type instance Item (S thunk) t = Subst (thunk ()) t
\end{spec}

\paragraph{Modifying unification.} Our goal is to ensure that when a
variable is unified and receives a concrete binding, any associated
thunks are immediately triggered. In \ToolName{}, we cannot directly modify
``the unification algorithm'' globally, since each type is free to
define its own unification logic. However, because the substitution
interface is tightly controlled, all modifications to the substitution
state must go through the public method:

\il{insertSubM :: (MonadState (AllSubs m) m, Typeable t) => ID -> t -> m ()}

\noindent This centralized entry point allows us to extend
substitution behavior in a principled way. For instance, we can modify
\il{insertSubM} to support deferred computations (or thunks) associated
with variable bindings. Although we omit the implementation details
here, the extended behavior proceeds as follows:

\begin{enumerate}
\item Retrieve any thunk associated with the given \il{ID}, if one exists.
\item Insert the new association into the substitution map.
\item If a thunk was retrieved, execute it with the new value associated to \il{ID}.
\end{enumerate}

\paragraph{Providing an interface to register new thunks.} We
extend the substitution interface with a method to add a new thunk:

\begin{spec}[frame=none]
insertThunkM :: (MonadState (AllSubs m) m, Typeable t)
             => Proxy t -> ID -> m () -> m ()}
\end{spec}

\noindent The \il{Proxy t} argument reifies the type \il{t} without needing
a value, allowing us to locate the correct substitution map. The thunk
(here some monad \il{m ()}) is stored and later executed when the
variable with the given \il{ID} is bound.

With all this, we can finally construct \il{lazy}, a \ToolName{}-level
primitive, that allows users to postpone the application of some
predicate to a logical term.

\begin{spec}
lazy :: forall t. Unifiable t => (t -> Goal ()) -> t -> Goal ()
foo `lazy` t = do
  tWalked <- getWalked t
  case isUVar tWalked of
    Just tid -> insertThunkM (Proxy::Proxy t) tid (foo `lazy` tWalked)
    Nothing -> foo tWalked
\end{spec}

\subsection{Disequality With Lazyness}
\label{sec:disequality}

Lazy disequality is based on an insight from
\texttt{faster-miniKanren}~\cite{faster-miniKanren}: a disequality
constraint can only fail when all of its component variables are fully
instantiated and equal. This means that disequality checks do not need
to be tracked or re-evaluated at every unification step. Instead,
evaluation can be deferred until sufficient information becomes
available—specifically, when the relevant variables are ground.

A downside of this delayed approach is that it can permit inconsistent
states, especially for small, finite types. For instance, the example
below results in a constraint that \texttt{b} must be neither
\il{True} nor \il{False}, which is clearly unsatisfiable, but the
system cannot detect this until further instantiation
occurs.

\begin{spec}
inconsistent :: Goal ()
inconsistent = do
  b <- fresh
  b =\= UT True
  b =\= UT False
\end{spec}

To address this issue, we implement a new class, \il{Disunifiable},
that allows every type to implement its own disequality strategy. For
example, \il{Boolean}s can eagerly check disequality structurally and
a sets might implement disequality wihout relying on its internal
representation.

We are acutely aware that introducing additional type classes runs against our
general focus on simplicity. In Haskell, type class proliferation can
make code harder to read and write by increasing the number of
constraints in type signatures, which in turn burdens both the
programmer and the type checker. It can also obscure intent through
overloading, where the behavior of a function depends on instances
that may not be locally visible or obvious. Despite these concerns, we
found that \il{Disunifiable} offers a clear and modular solution to a
real need, and its benefits outweigh the added complexity in this
case.

\begin{spec}
class Unifiable t => DisUnifiable t where
  disunify :: Goal () -> t -> t -> Goal ()

(=/=) :: DisUnifiable t => t -> t -> Goal ()
a =/= b = disunify gFail a b
\end{spec}

The definition of \il{disunify} includes a continuation, of type
\il{Goal ()}, that is invoked only when the two terms are equal up to
the current point of traversal. This continuation dictates how to
proceed with checking the remaining structure for disequality. If the
entire values turn out to be equal, the continuation ultimately fails
via \il{gFail}. Recursive calls to \il{disunify} are threaded through
the continuation, enabling lazy traversal and allowing the check to
terminate early as soon as a difference is encountered.

Straightforward instances of \il{DisUnifiable}, based on structural
equality, can be automatically derived using generics. For
illustration, however, we show a manual instance for pairs:

\begin{spec}
instance (Eq a, Eq b, DisUnifiable a, DisUnifiable b) =>
         DisUnifiable (U (a, b)) where
    disunify k x y | x == y = k
    disunify k x y@(UVar _) = lazy (disunify k x) y
    disunify k x@(UVar _) y = lazy (flip (disunify k) y) x
    disunify k (U (x1, y1)) (U (x2, y2)) =
      disunify (disunify k y1 y2) x1 x2
\end{spec}

This instance performs structural disequality by recursively
disunifying components. Lazy traversal is used for unbound variables
to avoid prematurely forcing computation.

%\section{Product Types}
%\section{Debugging}

\begin{ignore}
\begin{code}
{-# OPTIONS_GHC -w #-}
{-# LANGUAGE FlexibleInstances #-} -- REmove after isString is moved
{-# LANGUAGE TypeApplications #-}
{-# LANGUAGE OverloadedStrings #-}
{-# LANGUAGE DeriveGeneric #-}
module Usability where

import Walrus

import           Walrus hiding (VStr)
import           Prettyprinter
import           GHC.Generics (Generic)

-- Move this to the core library
import           Data.String

instance IsString (UT String) where
  fromString = UT
\end{code}
\end{ignore}

\section{Product Types}
\label{sec:productTypes}

Tuples and records are a foundational abstraction in modern
programming. They are widely used in functional languages like
Haskell, where pattern matching and field accessors provide
convenient, readable mechanisms for manipulation. In logic
programming, however, working with product types can be more
cumbersome. Accessing a single field in a record often requires
destructuring the entire value manually, introducing fresh variables
and explicit unification steps.

For example, consider how a field is accessed in a record type with 6 fields. In
Haskell, this is straightforward using either field accessors or
pattern matching:

\begin{ignore}
\begin{code}
data Person = Person
  { name     :: UT String
  , age      :: UT Int
  , address  :: UT String
  , email    :: UT String
  , phone    :: UT String
  , employed :: UT Bool
  }
  deriving (Eq, Show, Generic)

instance Pretty Person where
pretty = undefined
instance PreUnifiable Person
\end{code}
\end{ignore}

\begin{code}
-- Haskell: field accessor
calledCarlos :: Person -> Bool
calledCarlos person =
  let nm = name person in
    nm == "Carlos"

-- Haskell: pattern matching in a monadic context
calledCarlosM :: Monad m => m Person -> m Bool
calledCarlosM personM = do
  Person nm _ _ _ _ _ <- personM
  return $ nm == "Carlos"
\end{code}

In logic programming, however, the equivalent operation typically
requires explicit unification with fresh variables:

\begin{code}
-- explicit unification
calledCarlosG :: U Person -> Goal ()
calledCarlosG p = do
  (name, age, address) <- fresh3
  (email, phone, employed) <- fresh3
  p === U (Person name age address email phone employed)
  name === "Carlos"
\end{code}

In our experience developing with \ToolName{}, this pattern appeared
frequently: functions often began with repetitive boilerplate just to
extract individual fields. This kind of manual unpacking is verbose
and can introduce unnecessary variables, making the code harder to
read, reason about, and maintain. Feedback from early users of the
language highlighted this as a recurring pain point.

To address this, \ToolName{} introduces the \il{UnifiableProduct}
class, which supports generic construction and deconstruction of
product types. This enables cleaner, more idiomatic logic code that
resembles conventional functional programming. For instance, the
previous relation can be rewritten using either pattern matching or
lifted field accessors:

\begin{code}
instance UnifiableProduct Person where

-- Field accessors
calledCarlosFA :: U Person -> Goal ()
calledCarlosFA person = do
  pName <- wLift name person
  pName === "Carlos"

-- Pattern matching
calledCarlosPM :: U Person -> Goal ()
calledCarlosPM person = do
  Person nm _ _ _ _ _ <- patternProd person
  nm === "Carlos"
\end{code}

The core of this functionality is the \il{freshProd} method provided by
the \il{UnifiableProduct} class. It constructs a fresh instance of a
product type by generating fresh values for each of its
components. This produces a Haskell type that can be pattern matched and
traversed normally.

\begin{spec}
class PreUnifiable t => UnifiableProduct t where
  freshProd :: Goal t
\end{spec}

Building on this foundation, \ToolName{} provides two derived
utilities that mirror common functional programming idioms:

\begin{itemize}

\item \il{patternProd} enables pattern matching by ensuring the
product is instantiated via \il{freshProd} so that product types can
be destructured clearly and directly.

\item \il{wLift} lifts \emph{any} function over a product type into
the logic monad. Given \il{f :: p -> b}, it produces
\il{(wLift f) :: U p -> Goal b}.
This is particularly useful to lift field accessors
and tuple projections like \il{fst} and \il{snd}.

\end{itemize}

These facilities allow users to work with records and tuples in logic
code with the same natural syntax and convenience as in
Haskell. Moreover, both \il{wLift} and \il{patternProd} are optimized
to avoid unnecessary instantiations: they invoke \il{freshProd} only
when the input is a logical variable. If the argument is already a
concrete term, no additional variables are introduced. This
optimization reduces both the number of unifications performed and the
overall size of the substitution.

%%%%%%%%%%%%%%%%%%%%%%%%%%%%%%%%%%%%%%%%%%%%%%%%%%%%%%%%%%%%%%

\section{Debugging}
\label{sec:debug}

Debugging logic programs is notoriously challenging due to the
presence of interleaving, backtracking, and, in the case of
\ToolName{}, on-demand laziness. To support effective debugging,
\ToolName{} provides structured execution tracing as a built-in
feature. The core facility is the function \il{debugGoal :: Goal t ->
Int -> DB.Tree}, which runs a logic relation for a fixed number of
steps and collects a trace of the computation. The resulting trace
captures the structural shape of the computation as a tree of
execution steps, where branches represent nondeterministic choices. By
abstracting away from interleaving and evaluation order, the trace
makes execution easier to inspect, reason about, and debug.

\begin{figure}[h]
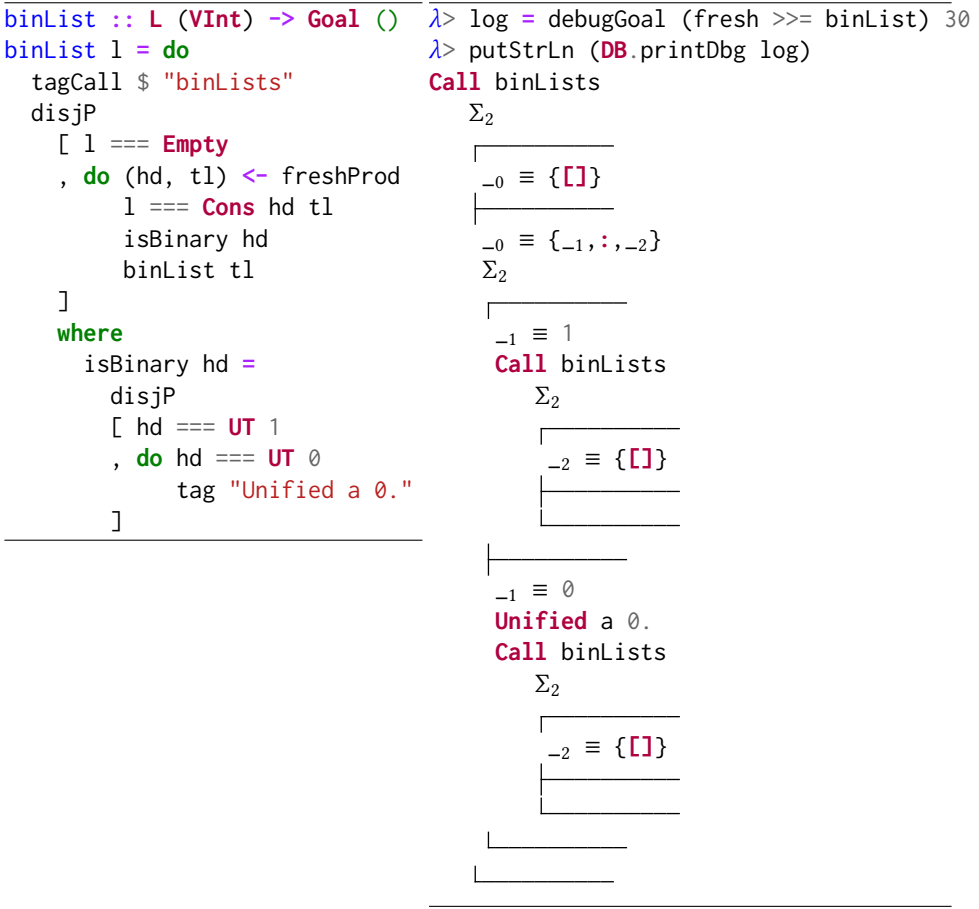


\begin{subfigure}[T]{.40\textwidth}

\begin{code}
binList :: L (VInt) -> Goal ()
binList l = do
  tagCall $ "binLists"
  disjP
    [ l === Empty
    , do (hd, tl) <- freshProd
         l === Cons hd tl
         isBinary hd
         binList tl
    ]
    where
      isBinary hd =
        disjP
        [ hd === UT 1
        , do hd === UT 0
             tag "Unified a 0."
        ]
\end{code}
\end{subfigure}
\begin{subfigure}[T]{.5\textwidth}
\begin{minted}[escapeinside=??,autogobble,frame=lines]{haskell}
λ> log = debugGoal (fresh >>= binList) 30
λ> putStrLn (DB.printDbg log)
Call binLists
   ?$\Sigma_2$?
   ?┌──────────?
    _₀ ?$\equiv$? {[]}
   ?├──────────?
    _₀ ?$\equiv$? {_₁,:,_₂}
    ?$\Sigma_2$?
    ?┌──────────?
     _₁ ?$\equiv$? 1
     Call binLists
        ?$\Sigma_2$?
        ?┌──────────?
         _₂ ?$\equiv$? {[]}
        ?├──────────?
        ?└──────────?
    ?├──────────?
     _₁ ?$\equiv$? 0
     Unified a 0.
     Call binLists
        ?$\Sigma_2$?
        ?┌──────────?
         _₂ ?$\equiv$? {[]}
        ?├──────────?
        ?└──────────?
    ?└──────────?
   ?└──────────?
\end{minted}

\end{subfigure}

\caption{Predicate identifying binary lists (left) and the trace
resulting from evaluating the relation for 30 steps (right). This
execution corresponds to idetifying the first three binary lists
(\il{[]}, \il{[1]}, and \il{[0]}) before halting.}

\label{fig:debug}

\end{figure}

To illustrate the tracing system, consider the predicate \il{binList}
in Figure~\ref{fig:debug}, which defines lists containing only 0s and
1s. The \il{tagCall} annotation marks the function call, and the
\il{tag} annotation inserts a custom message. The trace can be printed
as a compact symbolic notation to represent control flow as in
Figure~\ref{fig:debug}~(right). Function calls are shown with
\texttt{Call <name>} and branching points, produced by \il{disjP}, are
labeled with $\Sigma_n$, indicating a nondeterministic choice among $n$
branches. Each branch is visually enclosed by a box using characters
\texttt{┌─}, \texttt{├─} and \texttt{└─}. Unification steps are shown
using the symbol $\equiv$, while custom debug annotations (e.g.,
\il{tag "Unified a 0."}) appear verbatim. Some branches in
Figure~\ref{fig:debug}~(right) appear empty because execution was
halted before they took a single step.

Behind the scenes, generating a structured trace like this requires
coordinating among different executions. In \ToolName{}, this is made
possible by two stateful mechanisms:

\begin{itemize}

\item \emph{Local execution path}: Our local state
\il{WState} records the path of execution as a list of integers
representing the branch taken at each disjunction.

\item \emph{Global trace ledger}: An append-only global ledger
accumulates trace entries, each tagged with the corresponding
execution path at the time of recording.

\end{itemize}

Using these paths, the function \il{debugGoal} converts the ledger into
a complete execution tree in the correct structural order.

\ToolName{} provides utilities for exploring, traversing, and
displaying execution trees. The function \il{DB.printDbg} renders a
formatted view of the execution as in Figure~\ref{fig:debug}. Other
utilities include \il{DB.printDbgOnlyCalls}, which filters the tree to
show only tagged function calls, and \il{DB.subTree :: DB.Tree ->
DB.Path -> DB.Tree}, which extracts a subtree at a given execution
path, zooming in on specific segments of a trace. Users can define
their own utilities to query or visualize execution trees in custom
ways.

\section{Evaluation}
\label{sec:evaluation}

\begin{ignore}
\begin{code}
module Evaluation where
\end{code}
\end{ignore}

We have only recently begun developing a benchmark suite for
\ToolName{} to better understand its performance characteristics and
the impact of various design decisions and, as discussed in
Section~\ref{sec:future-work}, optimization efforts are ongoing.
In this section, we evaluate the current performance of
\ToolName{} using five benchmarks that have emerged as de facto
standards for assessing miniKanren implementations and that we have
adapted from the \texttt{typedKanren} project~\cite{typedKanren},
which itself builds upon infrastructure introduced in
\texttt{klogic}~\cite{klogic}.

The first two benchmarks compute exponentials and logarithms using a
binary implementation of arithmetic. Our implementation is slightly
slower than the state-of-the-art \texttt{faster-miniKanren} but
faster than \texttt{typedKanren}.

The next three benchmarks use a lambda calculus interpreter to compute
quines, twines, and thrines: lists of one, two, or three lambda terms
such that each term evaluates to the next in the list, cyclically. We
presented our implementation of this interpreter, and its use in
computing quines, in Section~\ref{sec:example}.

Our interpreter takes advantage of the type system, which is one of
\ToolName{}’s main strengths, to guide evaluation. Unlike other
implementations, we do not need to impose additional runtime
constraints such as \il{absent}$^o$ to prevent lambda constructors
from being treated as variables~\cite{miniKanren-quines}.  The type
system ensures correctness and reduces reliance on disequality
constraints, which remain a performance bottleneck in \ToolName{}. As
a result, our typed interpreter performs comparably to
\texttt{typedKanren} and even outperforms it in the thrines
benchmark~\cite{typedKanren}.  For completeness, we also implemented
an untyped Scheme-style interpreter in \ToolName{}, which behaves
similarly to the one implemented by \texttt{typedKanren}, but runs at
roughly half its speed.

We believe this slowdown is largely due to the overhead introduced by
our straightforward but unoptimized implementation of
disequality. Anecdotally, we observe that our laziness-based approach
can be slower than hand-crafted solutions, particularly when solving
disequality constraints. However, further research is needed to
determine whether this performance gap is inherent to the approach or
if it can be mitigated through improved representations or heuristics.

\begin{table}[]
\begin{tabular}{|l||r|r|r|r|r|}
\hline
                   & faster-miniKanren & typedKanren & \ToolName{} & \ToolName{} untyped \\
\hline \hline
$3^n$ / n=5        & $83.18$           & $525.3$     & $124$       & -     \\ \hline
$log_3 n$ / n=243  & $13.48$           & $60.52$     & $14.5$      & -     \\ \hline
N quines / N=100   & $162.6$           & $696.3$     & $157$       & $1103$     \\ \hline
N twines / N=15    & $143.0$           & $1010$      & $274$       & $2135$     \\ \hline
N thrines / N=2    & $256.0$           & $2475$      & $183$       & 4431     \\ \hline

\end{tabular}

\caption{ Execution times (in microseconds) for selected benchmark
programs across several relational programming languages.  All
benchmarks were run on a MacBook Pro with an Apple M2 chip and 32~GB
of memory.  }

\label{table:othertools}

\end{table}

\begin{table}[h]
\begin{tabular}{|l||r|r|r|r|}
\hline
                   & Baseline & no type-segregation & With debugging \\ \hline \hline
$3^n$ / n=5        & $124$    & $119$                & $838$     \\ \hline
$log_3 n$ / n=243  & $14.5$   & $14.0$               & $90.2$    \\ \hline
N quines / N=100   & $157$    & $153$                & $698.0$   \\ \hline
N twines / N=15    & $274$    & $280$                & $1184$    \\ \hline
N thrines / N=2    & $183$    & $176$                & $894$     \\ \hline

\end{tabular}

\caption{Execution times (in microseconds) for variants of
\ToolName{}, each with a different feature removed. This isolates the
performance impact of type-segregated substitutions and debugging
infrastructure. All benchmarks were run on a MacBook Pro with an Apple
M2 chip and 32~GB of memory.}

\label{table:ablation}

\end{table}

To assess the performance impact of specific features in \ToolName{},
we performed an ablation study comparing the full system against two
modified variants. The first variant disables type-segregated
substitutions, falling back to a single unified substitution map for
all logical variables. In the arithmetic benchmarks, which involve only
a single type, this change introduced a 4\% overhead. In
contrast, the interpreter benchmarks, which involve around three
types, partially offset this cost; the variant performed better on the
more complex tests, where type segregation provides clearer
benefits. As expected, examples with more types and longer
substitutions benefit most from type-aware substitution. The second
variant enables \ToolName{}’s structured debugging infrastructure. As
anticipated, the added tracing incurs a substantial performance cost,
slowing execution by more than a factor of four.

\section{Future Work}
\label{sec:future-work}

\begin{ignore}
\begin{code}
module FutureWork where
\end{code}
\end{ignore}

The miniKanren ecosystem offers such a rich and playful landscape that
it is difficult not to get excited about the future possibilities of
\ToolName{}. In that spirit, we hope the reader will indulge the
speculative nature of this section, where we outline directions that
are both ambitious and, we believe, within reach.

We have begun developing a benchmark suite for \ToolName{} to better
understand the performance of our implementation and identify key
bottlenecks. We plan to use these benchmarks to assess the
effectiveness of various optimization strategies, including:

\begin{itemize}

\item \textbf{Tabling}: Caching intermediate results to avoid
redundant computation across similar subgoals. \ToolName{}'s
\il{Goal} carries a global state which can be used for memoization.

\item \textbf{Path pruning}: Implementing early path
shortening strategies to reduce long \il{walk}s.

\item \textbf{In-memory state management}: Investigating whether
techniques used in \texttt{Molog}~\cite{Molog} offer meaningful performance gains
for \ToolName{}.

\item \textbf{Implement an equivalent to
\il{set-var-val!}}~\cite{faster-miniKanren}: While we wouldn't want to
modify variables directly in Haskell, we can use similar ideas to
reduce the size of the substitutions.

\item \textbf{Integration with external solvers}: Exploring the
feasibility and benefits of offloading subproblems to SMT solvers or
other backend systems during execution.

\item \textbf{Heuristic-guided search}: We have already had some
success with exploiting the commutativity of conjunction with simple
heuristics to optimize evaluation order. We want to continue
developing this idea.

\end{itemize}

We will continue improving the usability of \ToolName{}, guided by our
experience using it in real-world scenarios. In the short term, our
main priority is addressing a key limitation that remains unresolved:
error reporting. This is particularly pressing given our focus on
bidirectional compilers. Diagnosing compilation failures is inherently
difficult, even in the conventional unidirectional setting.  While
\ToolName{}’s structured debugging tools have been valuable for
understanding execution behavior, they are not sufficient for
identifying and explaining why a program fails.

In Section~\ref{sec:lazy}, we discussed both the power and the risks
of \il{lazy} evaluation. This feature remains largely untamed, and we
believe its full potential is yet to be realized. We have begun
experimenting with dynamic techniques to (over)approximate when a
state has become inconsistent, as well as strategies for resolving
thunks after evaluation completes. We plan to continue exploring these
directions to gain better control and insight into \il{lazy} behavior
in realtional languages.

\section{Conclusion}
\label{sec:conclusion}

\begin{ignore}
\begin{code}
module Conclusion where
\end{code}
\end{ignore}

\ToolName{} is a typed logic programming language embedded in Haskell,
designed to make relational programming practical, modular, and
ergonomic. By combining a type-safe interface with features like lazy
execution, product type support, and structured debugging, \ToolName{}
makes it easier to write clear, reusable, and efficient logic
code. Our goal is to support engineering that makes logic
programming viable and practical in larger, real-world systems.

\section{References}
\bibliography{references}

\end{document}